\documentclass[twoside]{IEEEtran}
\PassOptionsToPackage{ruled, linesnumbered}{algorithm2e}
\usepackage[latin9]{inputenc}
\usepackage{verbatim}
\usepackage{algorithm2e}
\usepackage{amsmath}
\usepackage{amssymb}
\usepackage{graphicx}

\makeatletter
\newtheorem{theorem}{Theorem}

\usepackage{empheq}

\usepackage{balance}
\allowdisplaybreaks
\usepackage{url}
\usepackage{pgfplots}
\usepackage{pgfplotstable}
\usepackage{tikz}
\usetikzlibrary{calc}
\usepackage{cite}
\usepackage{xcolor}
\usepackage{mathtools}
\definecolor{myblue}{rgb}{0.0, 0.5, 1.0}
\definecolor{myred}{rgb}{1.0, 0.13, 0.32}
\definecolor{mygreen}{rgb}{0.31, 0.78, 0.47}

\newcommand{\herm}{^{\mathsf{H}}}
\newcommand{\trans}{^{\mathsf{T}}}

\DeclareMathOperator{\tr}{\mathsf{tr}}
\DeclareMathOperator{\vecd}{\mathsf{vec_d}}

\DeclareMathOperator{\diag}{\mathsf{diag}}
\DeclareMathOperator{\maximize}{maximize}
\DeclareMathOperator{\st}{subject~to}
\DeclareMathOperator{\argmin}{argmin}

\makeatother

\begin{document}
\title{{\huge{A Low-Complexity Solution to Sum Rate Maximization for IRS-assisted SWIPT-MIMO Broadcasting}}}
\author{
Vaibhav Kumar\IEEEauthorrefmark{1}, Anastasios Papazafeiropoulos\IEEEauthorrefmark{2}\IEEEauthorrefmark{3}, Muhammad Fainan Hanif\IEEEauthorrefmark{4}, \\ Le-Nam Tran\IEEEauthorrefmark{1},
and Mark F. Flanagan\IEEEauthorrefmark{1}\\
\IEEEauthorblockA{\IEEEauthorrefmark{1}School of Electrical and Electronic Engineering,
University College Dublin, Belfield, Dublin 4, Ireland\\
\IEEEauthorrefmark{2}Communications and Intelligent Systems Research Group, University of Hertfordshire, Hatfield AL10 9AB, U. K.\\
\IEEEauthorrefmark{3} Interdisciplinary Centre for Security, Reliability and Trust (SnT), University of Luxembourg, Luxembourg \\
\IEEEauthorrefmark{4}Institute of Electrical, Electronics and Computer
Engineering, University of the Punjab, Lahore, Pakistan\\
Email: vaibhav.kumar@ieee.org, tapapazaf@gmail.com, mfh21@uclive.ac.nz, \\ nam.tran@ucd.ie, mark.flanagan@ieee.org}
\thanks{This work was supported by Science Foundation Ireland under Grant
17/CDA/4786, and also in part by the Irish Research Council under Grant
IRCLA/2017/209.}}
\maketitle
\begin{abstract}
This paper focuses on the fundamental problem of maximizing the achievable weighted sum rate (WSR) at information receivers (IRs) in an intelligent reflecting surface (IRS) assisted simultaneous wireless information and power transfer system under a multiple-input multiple-output (SWIPT-MIMO) setting, subject to a quality-of-service (QoS) constraint at the energy receivers (ERs). Notably, due to the coupling between the transmit precoding matrix and the passive beamforming vector in the QoS constraint, the formulated non-convex optimization problem is challenging to solve. We first decouple the design variables in the constraints following a penalty dual decomposition method, and then apply an alternating gradient projection algorithm to achieve a stationary solution to the reformulated optimization problem. The proposed algorithm nearly doubles the WSR compared to that achieved by a block-coordinate descent (BCD) based benchmark scheme. At the same time, the complexity of the proposed scheme grows linearly with the number of IRS elements while that of the benchmark scheme is proportional to the cube of the number of IRS elements.
\end{abstract}

\begin{IEEEkeywords}
Intelligent reflecting surface, MIMO, SWIPT, energy harvesting, penalty
dual decomposition. 
\end{IEEEkeywords}

\section{Introduction}

The advancement in meta-materials technology has led to the development of intelligent reflecting surfaces (IRSs) which is being foreseen as a groundbreaking hardware technology for beyond-fifth-generation (B5G) and sixth-generation (6G) wireless communications systems~\cite{22-WC-industrialIRS}. Recent research has shown promising advantages of IRSs to support energy-efficient high-speed communication while also supporting massive connectivity. In parallel, simultaneous wireless information and power transfer (SWIPT) is another appealing technology to cater to the energy requirements of low-powered Internet-of-Things (IoT) devices~\cite{21-JSTSP-overviewWPT,21-uWMag-SWIPT-IoT}. In recent years, a significant research effort has been made towards investigating the benefits of IRSs in SWIPT-aided wireless communications systems, especially to improve the power transfer efficiency and to increase the operational range of energy receivers (ERs)~\cite{22-Proc-IRS-SWIPT}. 

In this context, one of the early works on IRS-aided SWIPT considered the problem of maximizing the weighted received sum power at the ERs subject to a signal-to-interference-plus-noise ratio (SINR) constraint at the information receivers (IRs) in an IRS-assisted SWIPT multiple-input single-output (MISO) system~\cite{20-WCL-maxPower}. Similarly, the fundamental problem of weighted sum rate (WSR) maximization (at the IRs) in an IRS-assisted SWIPT multiple-input multiple-output (MIMO) system, subject to a minimum weighted sum harvested power constraint (at the ERs) was considered in~\cite{20-JSAC-maxWSR}.  It is important to note that the beamforming optimization problems in IRS-assisted systems are challenging to solve in general, due to the coupling of the design variables in the objective and/or constraint(s). Although alternating optimization (AO) based schemes are one of the most popular approaches to tackle such problems in IRS-assisted communications, a near-optimal solution is not guaranteed if the design variables are coupled in the constraints (see~\cite{23-WCL-SCA} and the references therein).

It is well-known that the problem of WSR maximization in a SWIPT-MIMO system is similar to that of WSR maximization in a MIMO system subject to one or more interference constraints (e.g., underlay spectrum sharing MIMO systems). Therefore, for the WSR maximization problem in the IRS-assisted SWIPT-MIMO system, the authors in~\cite{20-JSAC-maxWSR} followed the approach of WSR maximization proposed in~\cite{20-TVT-CR-IRS-MIMO} and~\cite{22-TWC-CR-IRS-MIMO}. In particular, to obtain the optimal transmit precoding matrices (TPMs) and the passive beamforming vector at the IRS that jointly maximize the WSR, a block-coordinate descent (BCD)
method was used in~\cite{20-JSAC-maxWSR}. It is interesting to note that the shortcomings (in terms of performance and computational complexity) of the BCD-based beamforming design approach for the IRS-assisted MIMO underlay spectrum sharing system were highlighted in~\cite{22-WCL-PDDGP}, where the authors also proposed a high-performance and low-complexity solution using a penalty dual decomposition based alternating gradient projection (PDDAGP) method. Motivated by this observation, in this paper we propose the PDDAGP method for optimal beamforming design in the IRS-aided SWIPT-MIMO system, which results in a significantly higher WSR than that achieved by the BCD-based approach, and also incurs a notably lower complexity compared to the benchmark scheme.

\paragraph*{Notations}

Bold uppercase and lowercase letters respectively denote matrices and vectors. For a complex-valued matrix $\mathbf{X}$, the (ordinary) transpose, conjugate transpose, trace, determinant, and Frobenius norm are denoted by $\mathbf{X}\trans$, $\mathbf{X}\herm$, $\tr(\mathbf{X})$, $|\mathbf{X}|$, and $\|\mathbf{X}\|$, respectively. The absolute value of a complex number $x$ is denoted by $|x|$. The vector space of all complex-valued matrices of size $M\times N$ is denoted by $\mathbb{C}^{M\times N}$. Using $\vecd(\mathbf{X})$ we denote a column vector formed from the elements on the main diagonal of $\mathbf{X}$. For a vector $\mathbf{x}$, $\diag(\mathbf{x})$ denotes a square diagonal matrix whose main diagonal has the same elements as those of $\mathbf{x}$. The complex-valued gradient of a function $f(\cdot)$ with respect to (w.r.t.) $\mathbf{X}^{*}$ is denoted by $\nabla_{\mathbf{X}}f(\cdot)$, where $\mathbf{X}^{*}$ represents the complex conjugate of $\mathbf{X}$, and Euclidean projection of $\mathbf{X}$ onto the set $\mathcal{X}$ is defined by $\Pi_{\mathcal{X}}\{\mathbf{x}\}\triangleq\argmin_{\hat{\mathbf{x}}\in\mathcal{X}}\|\mathbf{x}-\hat{\mathbf{x}}\|$. The expectation operation is denoted by $\mathbb{E}\{\cdot\}$. The identity and zero matrices are respectively represented by $\mathbf{I}$ and $\boldsymbol{0}$, and $\sqrt{-1}$ is represented by $\iota$. 

\section{System Model and Problem Formulation}

Similar to~\cite{20-JSAC-maxWSR}, we consider an IRS-assisted SWIPT-MIMO system consisting of one base station (BS), $M_{\mathrm{I}}$ IRs, $M_{\mathrm{E}}$ ERs, and one passive IRS. It is assumed that the BS is equipped with $N_{\mathrm{B}}$ antennas, each of the IRs and ERs are respectively equipped with $N_{\mathrm{I}}$ and $N_{\mathrm{E}}$ antennas, respectively, and the IRS consists of $N_{\mathrm{S}}$ reflecting elements. The set of indices for IRs and ERs are respectively denoted by $\mathcal{M}_{\mathrm{I}}\triangleq\{1,2,\ldots,M_{\mathrm{I}}\}$ and $\mathcal{M}_{\mathrm{E}}\triangleq\{1,2,\ldots,M_{\mathrm{E}}\}$. The channel matrices for BS-IRS, BS-$m^{\mathrm{th}}$ IR, BS-$\ell^{\mathrm{th}}$ ER, IRS-$m^{\mathrm{th}}$ IR, and IRS-$\ell^{\mathrm{th}}$ ER are respectively denoted by $\mathbf{H}_{\mathrm{S}}\in\mathbb{C}^{N_{\mathrm{S}}\times N_{\mathrm{B}}}$, $\mathbf{H}_{m\mathrm{I}}\in\mathbb{C}^{N_{\mathrm{I}}\times N_{\mathrm{B}}}$, $\mathbf{H}_{\ell\mathrm{E}}\in\mathbb{C}^{N_{\mathrm{E}}\times N_{\mathrm{B}}}$, $\mathbf{G}_{m\mathrm{I}}\in\mathbb{C}^{N_{\mathrm{I}}\times N_{\mathrm{S}}}$ and  $\mathbf{G}_{\ell\mathrm{E}}\in\mathbb{C}^{N_{\mathrm{E}}\times N_{\mathrm{S}}}$. The IRS passive beamforming vector is denoted by $\boldsymbol{\phi}=[\phi_{1},\phi_{2},\ldots,\phi_{N_{\mathrm{S}}}]\trans\in\mathbb{C}^{N_{\mathrm{S}}\times1}$, where $\phi_{n_{\mathrm{S}}}\triangleq\exp(\iota\theta_{n_{\mathrm{S}}})$ and $\theta_{n_{\mathrm{S}}}\in[0,2\pi),\forall n_{\mathrm{S}}\in\mathcal{N_{\mathrm{S}}\triangleq}\{1,2,\ldots,N_{\mathrm{S}}\}$.\footnote{Although different IRS reflection models have been proposed in the literature, the unit-modulus model is the most frequently used (see e.g.,~\cite{23-WCL-SCA,20-WCL-maxPower,20-JSAC-maxWSR,20-JSAC-minPow,20-TVT-CR-IRS-MIMO,22-TWC-CR-IRS-MIMO,22-WCL-PDDGP}).} We assume the availability of perfect instantaneous channel state information (CSI) at the BS for all of the wireless links.\footnote{Although CSI acquisition is a challenging task in IRS-assisted communication systems, consideration of imperfect CSI is beyond the scope of this paper. The results presented in this paper serve as theoretical upper bounds on the performance of a practical system with imperfect CSI.} We denote the signal vector transmitted from the BS is given by $\mathbf{w}=\sum\nolimits  _{m\in\mathcal{M_{\mathrm{I}}}}\mathbf{F}_{m}\mathbf{s}_{m},$ where $\mathbf{s}_{m}\in\mathbb{C}^{\min\{N_{\mathrm{B}},N_{\mathrm{I}}\}\times1}$ is the signal vector intended for the $m^{\mathrm{th}}$ IR, and $\mathbf{F}_{m}\in\mathbb{C}^{N_{\mathrm{B}}\times\min\{N_{\mathrm{B}}\times N_{\mathrm{I}}\}}$ is the transmit precoding matrix (TPM) corresponding to $\mathbf{s}_{m}$. We assume that $\mathbb{E}\{\mathbf{s}_{m}\mathbf{s}_{m}\herm\}=\mathbf{I}$ and $\mathbb{E}\{\mathbf{s}_{m}\mathbf{s}_{m'}\herm\}=\boldsymbol{0}$ $\forall m\neq m'\in\mathcal{M}_{\mathrm{I}}$. The signal vector received at the $m^{\mathrm{th}}$ IR is given by $\mathbf{y}_{m\mathrm{I}}=(\mathbf{H}_{m\mathrm{I}}+\mathbf{G}_{m\mathrm{I}}\boldsymbol{\Phi}\mathbf{H}_{\mathrm{S}})\mathbf{w}+\mathbf{n}_{m\mathrm{I}},$ where $\boldsymbol{\Phi}\triangleq\diag(\boldsymbol{\phi})$, and $\mathbf{n}_{m\mathrm{I}}\in\mathbb{C}^{N_{\mathrm{I}}\times1}\sim\mathcal{CN}(\boldsymbol{0},\sigma_{m\mathrm{I}}^{2}\mathbf{I})$ is the additive white Gaussian noise (AWGN) vector at the $m^{\mathrm{th}}$ IR. Similarly, the received signal vector at the $\ell^{\mathrm{th}}$ ER is given by $\mathbf{y}_{\ell\mathrm{E}}=(\mathbf{H}_{\ell\mathrm{E}}+\mathbf{G}_{\ell\mathrm{E}}\boldsymbol{\Phi}\mathbf{H}_{\mathrm{S}})\mathbf{w}+\mathbf{n}_{\ell\mathrm{E}},$ where $\mathbf{n}_{\ell\mathrm{E}}\in\mathbb{C}^{N_{\mathrm{E}}\times1}\sim\mathcal{CN} (\boldsymbol{0},\sigma_{\ell\mathrm{E}}^{2}\mathbf{I})$ is the AWGN vector at the $\ell^{\mathrm{th}}$ ER. For the rest of this paper, we consider $\sigma_{m\mathrm{I}}^{2}=\sigma_{\ell\mathrm{E}}^{2}=\sigma^{2},\forall m\in\mathcal{M}_{\mathrm{I}},\ell\in\mathcal{M}_{\mathrm{E}}$. Also, with a slight abuse of notation, we define $\mathbf{H}_{\mathrm{S}}\leftarrow\mathbf{H}_{\mathrm{S}}/\sigma$,
$\mathbf{H}_{m\mathrm{I}}\leftarrow\mathbf{H}_{m\mathrm{I}}/\sigma$ and $\mathbf{H}_{\ell\mathrm{E}}\leftarrow\mathbf{H}_{\ell\mathrm{E}}/\sigma$; this normalization step will mitigate potential numerical issues caused by dealing with extremely small values. We further define $\mathbf{Z}_{m}\triangleq\mathbf{H}_{m\mathrm{I}}+\mathbf{G}_{m\mathrm{I}}\boldsymbol{\Phi}\mathbf{H}_{\mathrm{S}}$ and $\boldsymbol{\Xi}_{\ell}\triangleq\mathbf{H}_{\ell\mathrm{E}}+\mathbf{G}_{\ell\mathrm{E}}\boldsymbol{\Phi}\mathbf{H}_{\mathrm{S}}$. Therefore, the instantaneous achievable rate at the $m^{\mathrm{th}}$ IR is given by 
\begin{equation}
R_{m}(\mathbf{X},\boldsymbol{\phi})=\ln\big|\mathbf{I}\!+\!\mathbf{Z}_{m}\mathbf{X}_{m}\mathbf{Z}_{m}\herm\mathbf{B}_{m}^{-1}\big|\!=\!\ln|\mathbf{A}_{m}|-\ln|\mathbf{B}_{m}|,\label{eq:mIR-Rate}
\end{equation}
where $\mathbf{X}\triangleq\{\mathbf{X}_{m}\}_{m\in\mathcal{M}_{\mathrm{I}}}$, $\mathbf{X}_{m}\triangleq\mathbf{F}_{m}\mathbf{F}_{m}\herm$ (this is the transmit covariance matrix), $\mathbf{A}_{m}\triangleq\mathbf{I}+\mathbf{Z}_{m}\boldsymbol{\Sigma}\mathbf{Z}_{m}\herm$, $\boldsymbol{\Sigma}\triangleq\sum_{k\in\mathcal{M}_{\mathrm{I}}}\mathbf{X}_{k}$, $\mathbf{B}_{m}\triangleq\mathbf{I}+\mathbf{Z}_{m}\boldsymbol{\Sigma}_{m}\mathbf{Z}_{m}\herm$ (this is the interference-plus-noise covariance matrix), and $\boldsymbol{\Sigma}_{m}\triangleq\boldsymbol{\Sigma}-\mathbf{X}_{m}$. The total  harvested power at the $\ell^{\mathrm{th}}$ ER is given by $P_{\ell\mathrm{H}} (\mathbf{X},\boldsymbol{\phi})=\eta\tr\big(\boldsymbol{\Xi}_{\ell}\boldsymbol{\Sigma}\boldsymbol{\Xi}_{\ell}\herm\big),$ where $0<\eta\leq1$ is the energy harvesting efficiency. Therefore, a WSR maximization problem for the IRS-assisted SWIPT-MIMO system can be formulated as follows: 
\begin{subequations}
\label{eq:OptProbOrig}
\begin{align}
\underset{\mathbf{X},\boldsymbol{\phi}}{\maximize}\  & \big\{ R_{\mathrm{sum}}\big(\mathbf{X},\boldsymbol{\phi}\big)\triangleq\sum\nolimits _{m\in\mathcal{M}_{\mathrm{I}}}\omega_{m}R_{m}(\mathbf{X},\boldsymbol{\phi})\big\}\label{eq:ObjOrig}\\
\st\  & P_{\mathrm H} \big( \mathbf X, \boldsymbol \phi \big) \geq 1,\label{eq:EHC}\\
 & \tr\big(\boldsymbol \Sigma \big)\leq P_{\mathrm{B}},\label{eq:TPC}\\
 & |\phi_{n_{\mathrm{S}}}|=1 \ \forall n_{\mathrm{S}}\in\mathcal{N}_{\mathrm{S}}.\label{eq:UMCs}
\end{align}
\end{subequations}
In~\eqref{eq:OptProbOrig}, $\omega_{m}$ denotes the rate weighting factor for the $m^{\mathrm{th}}$ IR, $P_{\mathrm{H}}\big(\mathbf{X},\boldsymbol{\phi}\big)\triangleq \big(1/\tilde{P}_{\mathrm{th}}\big)\sum_{\ell\in\mathcal{M}_{\mathrm{E}}}\alpha_{\ell}P_{\ell\mathrm{H}}$, $\tilde{P}_{\mathrm{th}}\triangleq P_{\mathrm{th}}/\sigma^{2}$ with $P_{\mathrm{th}}$ being the total minimum weighted power required to be harvested at the ERs, $\alpha_{\ell}$ is the harvested power weighting factor at the $\ell^{\mathrm{th}}$ ER, and $P_{\mathrm{B}}$ denotes the transmit power budget at the BS. It is easy to observe that~\eqref{eq:OptProbOrig} is non-convex due to the coupling of the design variables (i.e., $\mathbf{X}$ and $\boldsymbol{\phi}$) in~\eqref{eq:ObjOrig} and~\eqref{eq:EHC}, and the non-convex constraints in~\eqref{eq:UMCs}. 

In order to solve~\eqref{eq:OptProbOrig}, the authors in~\cite{20-JSAC-maxWSR} first reformulated the problem by using the conventional weighted minimum mean-square error (WMMSE) method and then used an alternating optimization (AO) based BCD approach. It will be seen in Sec.~\ref{sec:Results} that the BCD-based approach of~\cite{20-JSAC-maxWSR} results in a notably inferior performance.
Moreover, as discussed in~\cite[Sec.~III-D]{20-JSAC-maxWSR}, the complexity of the BCD method grows as $\mathcal{O}(N_{\mathrm{S}}^{3})$ for large-scale systems where $N_{\mathrm{S}}\gg\max\{N_{\mathrm{B}},N_{\mathrm{I}},N_{\mathrm{E}},M_{\mathrm{I}},M_{\mathrm{E}}\}$, which represents the practical case where a significant benefit is achieved by the IRS. Also, note that for channels with relatively low coherence time (such as rapidly varying fast-fading channels), it may be practically infeasible to run a high-complexity optimization process as frequently as is needed to update the optimal beamforming design.

\section{Proposed Solution}

To tackle the coupling between the design variables (i.e., $\mathbf{X}$ and $\boldsymbol{\phi}$), in the constraint~\eqref{eq:EHC}, we follow the method of penalty dual decomposition, originally proposed in~\cite{20-TSP-PDD}. For this purpose, we first define $f\big(\mathbf{X},\boldsymbol{\phi},\tau\big) \triangleq 1+\tau-P_{\mathrm{H}}\big(\mathbf{X},\boldsymbol{\phi}\big)$; note that for some suitable $\tau\geq0$, $f\big(\mathbf{X},\boldsymbol{\phi},\tau\big)=0$ is equivalent to~\eqref{eq:EHC}. We now construct an augmented Lagrangian objective function defined as 
\begin{align}
\mathcal{R}_{\mu,\rho}(\mathbf{X},\boldsymbol{\phi},\tau) & \triangleq R_{\mathrm{sum}}\big(\mathbf{X},\boldsymbol{\phi}\big)\nonumber \\
 & -\big\{\mu f(\mathbf{X},\boldsymbol{\phi},\tau)+(0.5/\rho)f^{2}(\mathbf{X},\boldsymbol{\phi},\tau)\big\},\label{eq:AugObjDef}
\end{align}
 where $\mu$ is the Lagrange multiplier corresponding to the constraint $f\big(\mathbf{X},\boldsymbol{\phi},\tau\big)=0$ and $\rho$ is the penalty parameter. Therefore, for a given $(\mu,\rho)$, an equivalent optimization problem can be formulated as~(c.f.~\cite{20-TSP-PDD})
\begin{equation}
\underset{\mathbf{X},\boldsymbol{\phi},\tau}{\maximize}\big\{\mathcal{R}_{\mu,\rho}(\mathbf{X},\boldsymbol{\phi},\tau)\big|\tau\geq0,\eqref{eq:TPC},\eqref{eq:UMCs}\big\}.\label{eq:OptProbTransform}
\end{equation}
It is important to note that the design variables are decoupled in the constraints in~\eqref{eq:OptProbTransform}, and the coupling 
exists only in the objective function, i.e., $\mathcal{R}_{\mu,\rho}(\mathbf{X},\boldsymbol{\phi},\tau)$. 

Before proposing a low-complexity and high-performance algorithm to obtain a stationary solution to~\eqref{eq:OptProbTransform}, we derive closed-form expressions for $\nabla_{\mathbf{X}}\mathcal{R}_{\mu,\rho}\big(\mathbf{X},\boldsymbol{\phi},\tau\big)$
and $\nabla_{\boldsymbol{\phi}}\mathcal{R}_{\mu,\rho}\big(\mathbf{X},\boldsymbol{\phi},\tau\big)$.
One can easily note that $\nabla_{\mathbf{X}}\mathcal{R}_{\mu,\rho}\big(\mathbf{X},\boldsymbol{\phi},\tau\big)=\big\{\nabla_{\mathbf{X}_{m}}\mathcal{R}_{\mu,\rho}\big(\mathbf{X},\boldsymbol{\phi},\tau\big)\big\}_{m\in\mathcal{M}_{\mathrm{I}}}$.
A closed-form expression for $\nabla_{\mathbf{X}_{m}}\mathcal{R}_{\mu,\rho}(\mathbf{X},\boldsymbol{\phi},\tau)$
is given in the following theorem.
\begin{theorem}
\label{thm:gradXClosed} A closed-form expression for $\nabla_{\mathbf{X}_{m}}\mathcal{R}_{\mu,\rho}\big(\mathbf{X},\boldsymbol{\phi},\tau\big)$
is given by $\nabla_{\mathbf{X}_{m}}\mathcal{R}_{\mu,\rho}\big(\mathbf{X},\boldsymbol{\phi},\tau\big)=\sum\nolimits _{k\in\mathcal{M}_{\mathrm{I}}}\omega_{k}\nabla_{\mathbf{X}_{m}}R_{k}\big(\mathbf{X},\boldsymbol{\phi}\big)+\big\{\mu+(1/\rho)f(\mathbf{X},\boldsymbol{\phi},\tau)\big\}\nabla_{\mathbf{X}_{m}}P_{\mathrm{H}}(\mathbf{X},\boldsymbol{\phi})$,
where 
\[
\nabla_{\mathbf{X}_{m}}R_{k}\big(\mathbf{X},\boldsymbol{\phi}\big)=\begin{cases}
\mathbf{Z}_{m}\herm\mathbf{B}_{m}^{-1/2}\text{\ensuremath{\mathbf{C}}}_{m}^{-1}\mathbf{B}_{m}^{-1/2}\mathbf{Z}_{m}, & \mathrm{if}\ m=k\\
\mathbf{Z}_{k}\herm\Big(\bar{\mathbf{B}}_{m,k}^{-1/2}\bar{\mathbf{C}}_{m,k}^{-1}\bar{\mathbf{B}}_{m,k}^{-1/2}\\
-\hat{\mathbf{B}}_{m,k}^{-1/2}\hat{\mathbf{C}}_{m,k}^{-1}\hat{\mathbf{B}}_{m,k}^{-1/2}\Big)\mathbf{Z}_{k}, & \mathrm{otherwise},
\end{cases}
\]
$\mathbf{C}_{m}\triangleq\mathbf{I}+\mathbf{B}_{m}^{-1/2}\mathbf{Z}_{m}\mathbf{X}_{m}\mathbf{Z}_{m}\herm\mathbf{B}_{m}^{-1/2}$,
$\bar{\mathbf{B}}_{m,k}\triangleq\mathbf{I}+\mathbf{Z}_{k}\boldsymbol{\Sigma}_{m}\mathbf{Z}_{k}\herm$,
$\bar{\mathbf{C}}_{m,k}\triangleq\mathbf{I}+\bar{\mathbf{B}}_{m,k}^{-1/2}\mathbf{Z}_{k}\mathbf{X}_{m}\mathbf{Z}_{k}\herm\bar{\mathbf{B}}_{m,k}^{-1/2}$,
$\hat{\mathbf{B}}_{m,k}\triangleq\mathbf{I}+\mathbf{Z}_{k}\boldsymbol{\Sigma}_{m,k}\mathbf{Z}_{k}\herm$,
$\boldsymbol{\Sigma}_{m,k}\triangleq\boldsymbol{\Sigma}_{m}-\mathbf{X}_{k}$,
$\hat{\mathbf{C}}_{m,k}\triangleq\mathbf{I}+\hat{\mathbf{B}}_{m,k}^{-1/2}\mathbf{Z}_{k}\mathbf{X}_{m}\mathbf{Z}_{k}\herm\hat{\mathbf{B}}_{m,k}^{-1/2}$,
and $\nabla_{\mathbf{X}_{m}}P_{\mathrm{H}}(\mathbf{X},\boldsymbol{\phi})=(\eta/\tilde{P}_{\mathrm{th}})\sum\nolimits _{\ell\in\mathcal{M}_{\mathrm{E}}}\alpha_{\ell}\boldsymbol{\Xi}_{\ell}\herm\boldsymbol{\Xi}_{\ell}$.
\end{theorem}
\begin{IEEEproof}
See Appendix~\ref{sec:proofGradX}.
\end{IEEEproof}
Next, we obtain a closed-form expression for the complex-valued gradient of $\mathcal{R}_{\mu,\rho} (\mathbf{X},\boldsymbol{\phi},\tau)$ w.r.t. $\boldsymbol{\phi}$. 
\begin{theorem}
\label{thm:gradThetaClosed}A closed-form expression for $\nabla_{\boldsymbol{\phi}}\mathcal{R}_{\mu,\rho}(\mathbf{X},\boldsymbol{\phi},\tau)$ is given by $\nabla_{\boldsymbol{\phi}}\mathcal{R}_{\mu,\rho}(\mathbf{X},\boldsymbol{\phi},\tau)=\sum\nolimits _{m\in\mathcal{M}_{\mathrm{I}}}\omega_{m}\nabla_{\boldsymbol{\phi}}R_{m}\big(\mathbf{X},\boldsymbol{\phi}\big)+\big\{\mu+(1/\rho)f(\mathbf{X},\boldsymbol{\phi},\tau)\big\}\nabla_{\boldsymbol{\phi}}P_{\mathrm{H}}\big(\mathbf{X},\boldsymbol{\phi}\big),$
where $\nabla_{\boldsymbol{\phi}}R_{m}\big(\mathbf{X},\boldsymbol{\phi}\big)=\vecd\big\{\mathbf{G}_{m\mathrm{I}}\herm\mathbf{D}_{m}\mathbf{H}_{\mathrm{S}}\herm\big\}$,
$\mathbf{D}_{m}\triangleq\mathbf{A}_{m}^{-1}\mathbf{Z}_{m}\boldsymbol{\Sigma}-\mathbf{B}_{m}^{-1}\mathbf{Z}_{m}\boldsymbol{\Sigma}_{m}$,
and $\nabla_{\boldsymbol{\phi}}P_{\mathrm{H}}\big(\mathbf{X},\boldsymbol{\phi}\big)=(\eta/\tilde{P}_{\mathrm{th}})\sum\nolimits _{\ell\in\mathcal{M}_{\mathrm{E}}}\alpha_{\ell}\vecd(\mathbf{G}_{\mathrm{\ell}\mathrm{E}}\herm\boldsymbol{\Xi}_{\ell}\boldsymbol{\Sigma}\mathbf{H}_{\mathrm{S}}\herm)$. 
\end{theorem}
\begin{IEEEproof}
See Appendix~\ref{sec:proofGradTheta}.
\end{IEEEproof}
We now propose the PDDAGP algorithm, shown in~\textbf{Algorithm~\ref{algoPDDGP}}, to attain a high-performance solution to~\eqref{eq:OptProbTransform}. We define $\mathcal{X}\triangleq\big\{\big\{\mathbf{X}_{m}\big\}_{m\in\mathcal{M}_{\mathrm{I}}}\big|\eqref{eq:TPC}\big\}$ as the feasible set of transmit covariance matrices $\mathbf{X}$. Similarly, $\varTheta\triangleq\big\{\boldsymbol{\phi}\big|\eqref{eq:UMCs}\big\}$ is defined as the feasible set for the design variable $\boldsymbol{\phi}$. In~\textbf{Algorithm~\ref{algoPDDGP}}, we use AO to iteratively update the variables $\mathbf{X}$ and $\boldsymbol{\phi}$. In the~$r^{\mathrm{th}}$ iteration, to update $\mathbf{X}^{(r)}$ for a given $\boldsymbol{\phi}^{(r)}$, we ascend in the direction of $\nabla_{\mathbf{X}}\mathcal{R}_{\mu,\rho}\big(\mathbf{X}^{(r)},\boldsymbol{\phi}^{(r)},\tau^{(r)}\big)$ with step size $\delta_{\mathbf{X}}$, and then project the resulting point onto the set $\mathcal{X}$ (see lines~4 and~5). After obtaining $\mathbf{X}^{(r+1)}$, we update $\boldsymbol{\phi}^{(r)}$ by ascending in the direction of  $\nabla_{\boldsymbol{\phi}}\mathcal{R}_{\mu,\rho}\big(\mathbf{X}^{(r+1)},\boldsymbol{\phi}^{(r)},\tau^{(r)}\big)$ using the step size $\delta_{\boldsymbol{\phi}}$, and then project the resultant vector onto $\varTheta$ to obtain $\boldsymbol{\phi}^{(r+1)}$
(lines~6 and~7). Next, following the constraint $\tau\geq0$ in~\eqref{eq:OptProbTransform}, we obtain $\tau^{(r+1)}$ as shown in line~8. The inner loop in~\textbf{Algorithm~\ref{algoPDDGP}} converges when $\Big[\mathcal{R}_{\mu,\rho}\big(\mathbf{X}^{(r+1)},\boldsymbol{\phi}^{(r+1)},\tau^{(r+1)}\big)-\mathcal{R}_{\mu,\rho}\big(\mathbf{X}^{(r)},\boldsymbol{\phi}^{(r)},\tau^{(r)}\big)\Big]/\mathcal{R}_{\mu,\rho}\big(\mathbf{X}^{(r)},\boldsymbol{\phi}^{(r)},\tau^{(r)}\big) < \epsilon$. Once the inner loop achieves convergence, we update the Lagrange multiplier $\mu$ and penalty multiplier $\rho$ as given in lines~10 and~11, respectively, and repeat the entire process. The outer loop converges when $\Big[\mathcal{R}_{\mu,\rho}\big(\mathbf{X}^{(r+1)},\boldsymbol{\phi}^{(r+1)},\tau^{(r+1)}\big)-R_{\mathrm{sum}}\big(\mathbf{X}^{(r+1)},\boldsymbol{\phi}^{(r+1)}\big)\Big]/\mathcal{R}_{\mu,\rho}\big(\mathbf{X}^{(r+1)},\boldsymbol{\phi}^{(r+1)},\tau^{(r+1)}\big)<\epsilon$. Note that projection onto $\mathcal{X}$ follows the standard water-filling solution, and projection onto $\varTheta$ is a simple scaling operation (see~\cite[eqn.~(6)]{22-WCL-PDDGP} for details). Moreover, appropriate values of $\delta_{\mathbf{X}}$ and $\delta_{\boldsymbol{\phi}}$ can be obtained using the backtracking line search scheme discussed in~\cite[eqn.~(8)]{22-WCL-PDDGP}. Following the arguments in~\cite{20-TSP-PDD}, it can be proved that when convergence is achieved, the stationary solution  of~\eqref{eq:OptProbTransform} becomes a stationary solution to~\eqref{eq:OptProbOrig}. Moreover, the convergence of the proposed PDDAGP algorithm can be readily proved following the line of argument in~\cite[Sec.~III-C]{22-WCL-PDDGP}, which is omitted here due to the space limitation. 

It is important to note that due to the contending constraints in~\eqref{eq:EHC} and~\eqref{eq:TPC}, the problem  in~\eqref{eq:OptProbOrig} may not be feasible for a given set of channels. If the outer loop in~\textbf{Algorithm~\ref{algoPDDGP}} does not converge within a certain number of iterations, we consider the problem to be \textit{infeasible} for that given set of channels. It is also noteworthy that the BCD-based solution proposed in~\cite[Algorithm~5]{20-JSAC-maxWSR} requires a \textit{feasible} TPM and $\boldsymbol{\phi}$ as initial points, whereas our proposed PDDAGP algorithm does not require feasible points for initialization. 

\begin{algorithm}[t]
\caption{The Proposed PDDAGP Algorithm to Solve~\eqref{eq:OptProbTransform}.}

\label{algoPDDGP}\small{

\KwIn{ $\mathbf{X}^{(0)}$, $\boldsymbol{\phi}^{(0)}$, $\tau^{(0)}$,
$\mu$, $\rho$, $\delta_{\mathbf{X}}$, $\delta_{\boldsymbol{\phi}}$,
$0<\kappa<1$}

\KwOut{ $\mathbf{X}^{\star}$, $\boldsymbol{\phi}^{\star}$}

\Repeat{convergence }{

$r\leftarrow1$

\Repeat{convergence }{

\tcc{Update $\mathbf{X}$}

$\!\!\!\!$Obtain $\nabla_{\mathbf{X}}\mathcal{R}_{\mu,\rho}\big(\mathbf{X}^{(r)},\boldsymbol{\phi}^{(r)},\tau^{(r)}\big)$
using Theorem~\ref{thm:gradXClosed}\;

$\!\!\!\!\mathbf{X}^{(r+1)}\!=\!\Pi_{\mathcal{X}}\big\{\mathbf{X}^{(r)}\!+\!\delta_{\mathbf{X}}\nabla_{\mathbf{X}}\mathcal{R}_{\mu,\rho}\big(\mathbf{X}^{(r)},\boldsymbol{\phi}^{(r)},\tau^{(r)}\big)\big\}$\;

\tcc{Update $\boldsymbol{\phi}$}

$\!\!\!\!$Obtain $\nabla_{\boldsymbol{\phi}}\mathcal{R}_{\mu,\rho}\big(\mathbf{X}^{(r+1)},\boldsymbol{\phi}^{(r)},\tau^{(r)}\!\big)$
using Theorem~\ref{thm:gradThetaClosed}\;

$\!\!\!\!\boldsymbol{\phi}^{(r+1)}\!\!=\!\!\Pi_{\varTheta}\big\{\boldsymbol{\phi}^{(r)}\!+\!\delta_{\boldsymbol{\phi}}\nabla_{\boldsymbol{\phi}}\mathcal{R}_{\mu,\rho}\big(\mathbf{X}^{(r+1)},\boldsymbol{\phi}^{(r)},\tau^{(r)}\big)\big\}$\;

\tcc{Update $\tau$}

$\!\!\!\!\tau^{(r+1)}=\max\{0,P_{\mathrm{H}}\big(\mathbf{X}^{(r+1)},\boldsymbol{\phi}^{(r+1)}\big)-1 -\mu \rho \}$\;

}

$\mu\leftarrow\mu+\frac{1}{\rho}f\big(\mathbf{X}^{(r+1)},\boldsymbol{\phi}^{(r+1)},\tau^{(r+1)}\big)$\;

$\rho\leftarrow\kappa\rho$\;

}

$\mathbf{X}^{\star}\leftarrow\mathbf{X}^{(r+1)}$, $\boldsymbol{\phi}^{\star}\leftarrow\boldsymbol{\phi}^{(r+1)}$,
$\tau^{\star}\leftarrow\tau^{(r+1)}$\;}
\end{algorithm}

\begin{figure}[t]
\begin{centering}
\includegraphics[width=0.7\columnwidth]{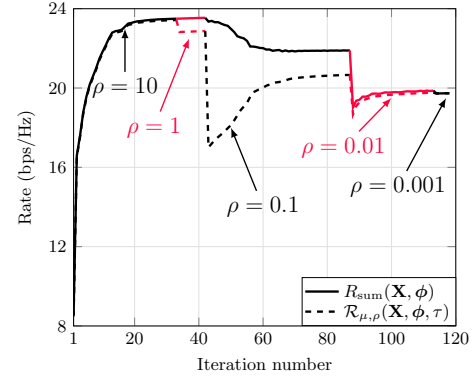}
\par\end{centering}
\caption{Convergence result of the proposed PDDGP algorithm at $P_{\mathrm{B}}=30$~dBm.}
\label{fig:convergence}
\end{figure}

\begin{figure*}[htp]
    \begin{minipage}[b]{0.32\linewidth}
        \centering
        \includegraphics[width=1\columnwidth]{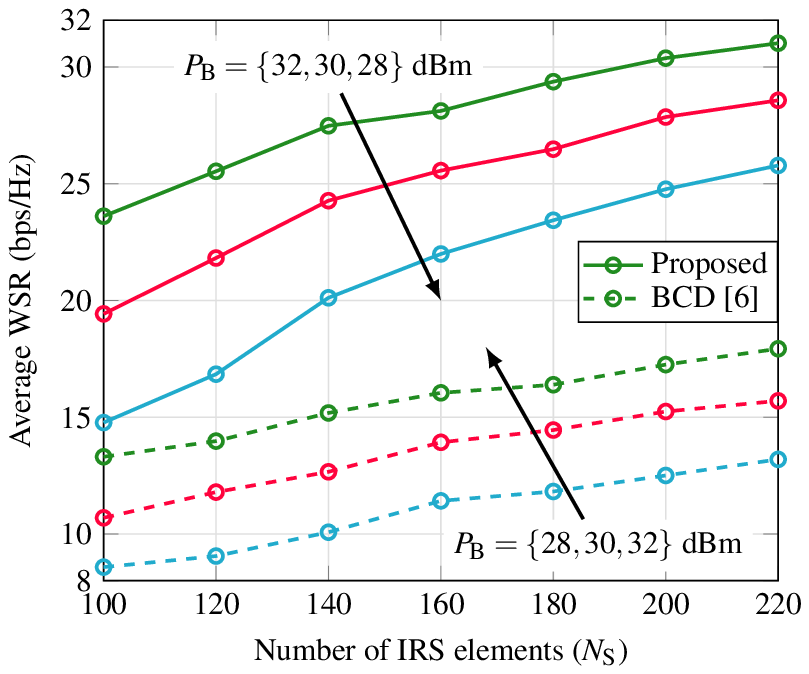}
        \caption{Impact of $N_{\mathrm S}$ on the average WSR.}
        \label{fig:rateN}
    \end{minipage}
  \hfill 
    \begin{minipage}[b]{0.32\linewidth}
        \centering
        \includegraphics[width=1\columnwidth]{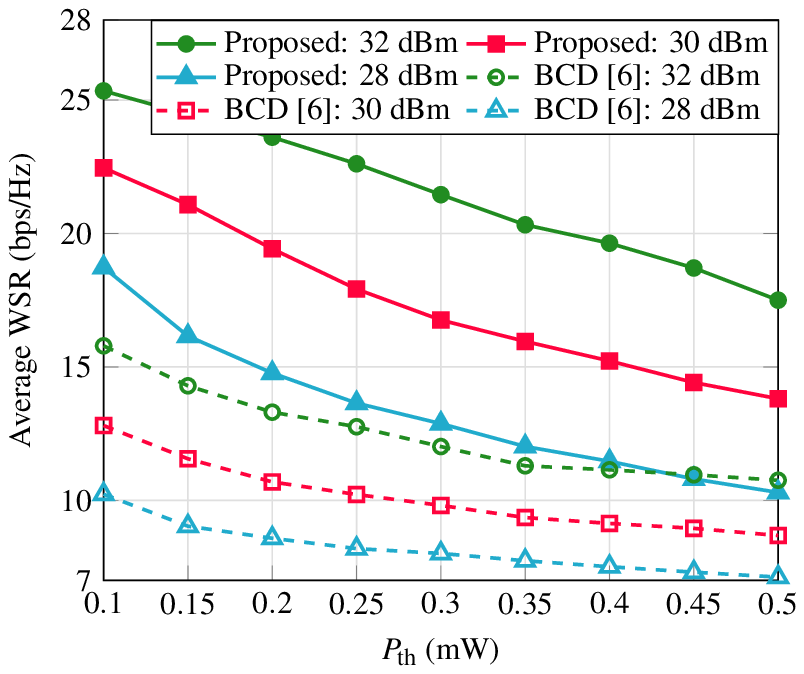}
        \caption{Effect of $P_{\mathrm{th}}$ on the average WSR.}
        \label{fig:ratePth}
    \end{minipage}
  \hfill 
    \begin{minipage}[b]{0.32\linewidth}
        \centering
        \includegraphics[width=1\columnwidth]{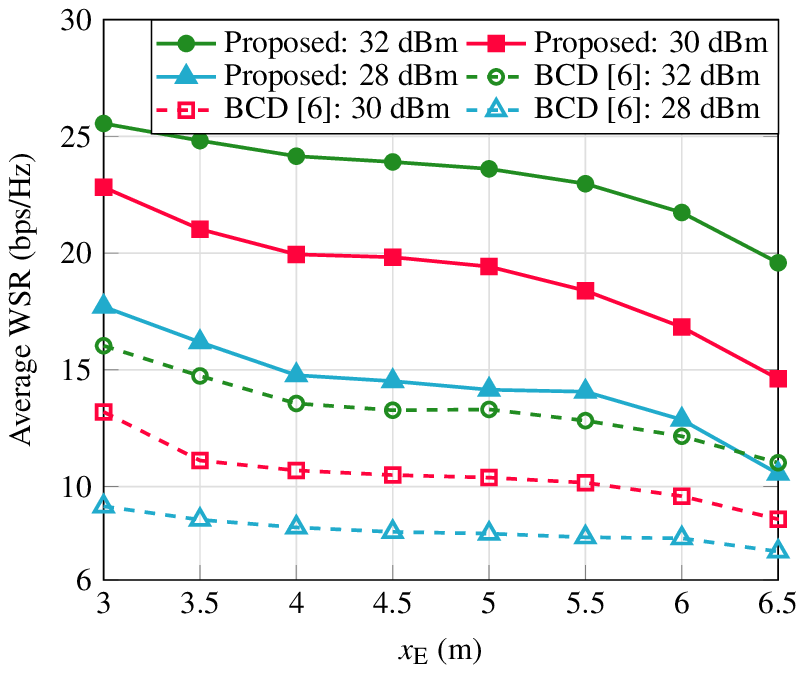}
        \caption{Effect of $x_{\mathrm{E}}$ on the average WSR.}
        \label{fig:rateXe}
    \end{minipage}
\end{figure*}




\paragraph*{Complexity Analysis}

Note that the complexity of the proposed PDDAGP algorithm is dominated by that of the inner loop in~\textbf{Algorithm~\ref{algoPDDGP}}. Defining the computational complexity as the number of complex-valued multiplications, it can be noted that the complexity of computing $\mathbf{Z}_{m},\forall m\in\mathcal{M}_{\mathrm{I}}$ and $\boldsymbol{\Xi}_{\ell},\forall\ell\in\mathcal{M}_{\mathrm{E}}$ are respectively given by $\mathcal{O}\big(M_{\mathrm{I}}N_{\mathrm{B}}N_{\mathrm{S}}\big(1+N_{\mathrm{I}}\big)\big)$ and $\mathcal{O}\big(M_{\mathrm{E}}N_{\mathrm{B}}N_{\mathrm{S}}N_{\mathrm{E}}\big)$. Similarly, given $\mathbf{X}^{(r)}$, $\boldsymbol{\phi}^{(r)}$, $\tau^{(r)}$, $\mathbf{Z}_{m},\forall m\in\mathcal{M}_{\mathrm{I}}$ and $\boldsymbol{\Xi}_{\ell},\forall\ell\in\mathcal{M}_{\mathrm{E}}$, the complexity of obtaining $\mathbf{X}^{(r+1)}$ is given by $\mathcal{O}\big(M_{\mathrm{E}}N_{\mathrm{E}}^{3}+M_{\mathrm{I}}^{3}N_{\mathrm{B}}^{3}+2M_{\mathrm{I}}^{2}N_{\mathrm{B}}^{2}N_{\mathrm{I}}+2M_{\mathrm{I}}^{2}N_{\mathrm{B}}N_{\mathrm{I}}^{2}+12M_{\mathrm{I}}^{2}N_{\mathrm{I}}^{3}+2M_{\mathrm{I}}N_{\mathrm{B}}N_{\mathrm{I}}^{2}-8M_{\mathrm{I}}N_{\mathrm{I}}^{3}\big)$. Analogously, the computational complexity of obtaining $\boldsymbol{\phi}^{(r+1)}$ is given by $\mathcal{O}\big(M_{\mathrm{I}}\big\{ N_{\mathrm{I}}^{3}+2N_{\mathrm{I}}N_{\mathrm{B}}\big(N_{\mathrm{B}}+N_{\mathrm{I}}\big)+N_{\mathrm{I}}N_{\mathrm{B}}N_{\mathrm{S}}+N_{\mathrm{I}}N_{\mathrm{S}}\big\}+M_{\mathrm{E}}\big(N_{\mathrm{B}}^{2}N_{\mathrm{S}}+N_{\mathrm{E}}N_{\mathrm{B}}N_{\mathrm{S}}+N_{\mathrm{S}}N_{\mathrm{E}}\big)\big)$. Note that since the complexity of projection operations will be comparatively smaller, we have neglected the associated terms. In the end, the complexity of computing $\tau^{(r+1)}$ is $\mathcal{O}\big(M_{\mathrm{E}}N_{\mathrm{B}}N_{\mathrm{E}}\big(N_{\mathrm{S}}+N_{\mathrm{B}}+N_{\mathrm{E}}\big)\big)$.
Therefore, the overall per-iteration complexity of~\textbf{Algorithm~\ref{algoPDDGP}} is given by $\mathcal{O}\big(M_{\mathrm{E}}N_{\mathrm{B}}\big(N_{\mathrm{B}}N_{\mathrm{E}}N_{\mathrm{S}}+N_{\mathrm{B}}N_{\mathrm{S}}+N_{\mathrm{E}}^{2}+2N_{\mathrm{E}}N_{\mathrm{S}}\big)+M_{\mathrm{E}}N_{\mathrm{E}}\big(N_{\mathrm{E}}^{2}+N_{\mathrm{S}}\big)+M_{\mathrm{I}}^{3}N_{\mathrm{B}}^{3}+M_{\mathrm{I}}N_{\mathrm{I}}N_{\mathrm{B}}\big(2M_{\mathrm{I}}N_{\mathrm{B}}+2M_{\mathrm{I}}N_{\mathrm{I}}+2M_{\mathrm{I}}N_{\mathrm{B}}+4N_{\mathrm{I}}+2N_{\mathrm{S}}\big)+12M_{\mathrm{I}}^{2}N_{\mathrm{I}}^{3}+M_{\mathrm{I}}\big(N_{\mathrm{B}}N_{\mathrm{S}}-7N_{\mathrm{I}}^{3}+N_{\mathrm{I}}N_{\mathrm{S}}\big)\big)$.
Since a practical deployment of an IRS-aided communication system is expected to involve a very large number of reflecting elements, it is expected that $N_{\mathrm{S}}\gg\max\big\{ N_{\mathrm{B}},N_{\mathrm{I}},N_{\mathrm{E}},M_{\mathrm{I}},M_{\mathrm{E}}\big\}$, and therefore, the per-iteration complexity of~\textbf{Algorithm~\ref{algoPDDGP}} is well approximated by $\mathcal{O}\big(N_{\mathrm{S}}\big(M_{\mathrm{E}}N_{\mathrm{E}}N_{\mathrm{B}}\big(2+N_{\mathrm{B}}\big)+2M_{\mathrm{I}}N_{\mathrm{I}}N_{\mathrm{B}}\big)\big)$, which is \emph{linear} in $N_{\mathrm{S}}$. This establishes the fact that the proposed PDDAGP algorithm is much more suitable for large-scale IRS-assisted SWIPT-MIMO systems in rapidly changing environments, compared to the BCD-based algorithm in~\cite{20-JSAC-maxWSR} whose complexity grows with the \emph{third power} of $N_{\mathrm{S}}$. 

\section{Numerical Results and Discussion} \label{sec:Results}

In this section, we present numerical results to establish the performance superiority of the proposed PDDAGP algorithm over the BCD-based scheme of~\cite{20-JSAC-maxWSR}. Similar to~\cite{20-JSAC-maxWSR}, we consider that the BS is located at $(0\ \mathrm{m},0\ \mathrm{m})$, the IRS is located at $(5\ \mathrm{m},2\ \mathrm{m})$, the IRs are uniformly and randomly distributed inside a circle of radius $4\ \mathrm{m}$ centered at $(400\ \mathrm{m},0\ \mathrm{m})$, and the ERs are uniformly and randomly distributed inside a circle of radius $1\ \mathrm{m}$ centered at $(x_{\mathrm{E}},0\ \mathrm{m})$. The path loss and small-scale fading models for all of the wireless links also follow~\cite{20-JSAC-maxWSR}. Furthermore, we assume $P_{\mathrm{th}}=0.2$~mW, $M_{\mathrm{I}}=2$, $M_{\mathrm{E}}=4$, $\omega_{m}=1 \ \forall m\in\mathcal{M}_{\mathrm{I}}$, $\alpha_{\ell}=1 \ \forall\ell\in\mathcal{M}_{\mathrm{E}}$, $\eta=0.5$, $N_{\mathrm{B}}=4$, $N_{\mathrm{I}}=N_{\mathrm{E}}=2$, $N_{\mathrm{S}}=100$, $x_{\mathrm{E}}=5\ \mathrm{m}$, $\kappa=0.1$, $\epsilon=10^{-3}$, a noise power spectral density of -160~dBm/Hz, and a total channel bandwidth of 1~MHz, \textit{unless stated otherwise}. The initial values are set as $\tau^{(0)} = 0$, $\mu^{(0)} = 0$, $\rho^{(0)} = 0$, $\mathbf X^{(0)} = \boldsymbol{0}$, and $\boldsymbol{\theta}^{(0)} = [1, 1, \ldots, 1]\trans$. In Figs.~\ref{fig:rateN}--\ref{fig:rateXe}, the average WSR is obtained over 100 random locations and independent small-scale fading realizations. Moreover, the numbers (in dBm) in the legends of Figs.~\ref{fig:ratePth} and~\ref{fig:rateXe} correspond to the value of $P_{\mathrm B}$.

Fig.~\ref{fig:convergence} shows a representative sample convergence result for the proposed PDDAGP algorithm, where each iteration corresponds to lines~3--9 in~\textbf{Algorithm~\ref{algoPDDGP}}. Following the arguments in~\cite[Sec.~III-C]{22-WCL-PDDGP}, it can be proved
that for a given $(\mu,\rho)$,~\textbf{Algorithm~\ref{algoPDDGP}} generates a strictly non-decreasing sequence of $\mathcal{R}_{\mu,\rho}\big(\mathbf{X},\boldsymbol{\phi},\tau\big)$. This fact is also evident in the figure. Once the inner loop in~\textbf{Algorithm~\ref{algoPDDGP} }converges, we update the Lagrange multiplier $\mu$ and decrease the value of the penalty parameter $\rho$. Due to a stricter penalty, $\mathcal{R}_{\mu,\rho}\big(\mathbf{X},\boldsymbol{\phi},\tau\big)$ drops suddenly (as seen in the figure when $\rho$ changes) and then for the new $(\mu,\rho)$, the sequence $\mathcal{R}_{\mu,\rho}\big(\mathbf{X},\boldsymbol{\phi},\tau\big)$ increases again. This whole process is repeated until the constraints in~\eqref{eq:EHC}--\eqref{eq:UMCs} are satisfied, which in turn nullifies the impact of the penalty in~\eqref{eq:AugObjDef}, resulting in the convergence of the algorithm. 

The impact of the number of IRS elements on the average WSR is shown in Fig.~\ref{fig:rateN}. With an increased number of elements, the IRS creates highly directed beams toward IRs and ERs, which results in an increase in the WSR. However, in contrast to the BCD-based approach of~\cite{20-JSAC-maxWSR}, the proposed algorithm enjoys the following benefits: (i) relaxed constraints in the latter (since the constraint in~\eqref{eq:EHC} is included in the objective in~\eqref{eq:OptProbTransform}), and (ii) the design variables $\mathbf{X}$ and $\boldsymbol{\phi}$ are decoupled in the constraints in the proposed algorithm. These benefits result in a larger beamforming gain compared to the BCD-based approach.
For the particular setting in this paper, the beamforming gain of the proposed PDDAGP-based method nearly doubles the average WSR compared to that achieved via the BCD-based approach. 

In Fig.~\ref{fig:ratePth}, we show the effect of increasing the value of weighted harvested power requirement ($P_{\mathrm{th}}$) on the average WSR for the proposed PDDAGP-based algorithm, and compare its performance with that of the BCD-based algorithm proposed in~\cite{20-JSAC-maxWSR}. As the value of $P_{\mathrm{th}}$ increases, the QoS constraints at the ERs become more demanding. This calls for a significant part of the beams from the BS and the IRS to be steered toward the ERs, resulting in a decrease in the WSR at the IRs. However, due to the luxury of relaxed constraints and decoupled optimization variables, the proposed PDDAGP algorithm results in superior beamforming designs compared to the BCD-based algorithm.

Fig.~\ref{fig:rateXe} shows the effect of the location of ERs on the WSR of the IRs. As the value of $x_{\mathrm{E}}$ increases (which increases the distance between the BS and ERs), the average channel quality of the BS-ER and IRS-ER links degrades, resulting in a challenging QoS constraint at the ERs. This in turn results in a decreased WSR at the IRs due to reasons similar to those discussed in the preceding paragraph. However, the proposed PDDAGP algorithm significantly outperforms the BCD-based benchmark solution. This also indicates that for given $P_{\mathrm{th}}$ and $R_{\mathrm{sum}}(\mathbf{X},\boldsymbol{\theta})$, the proposed algorithm helps to increase the operating distance of the ERs, i.e., it allows the ERs to be located further from the BS, compared to that facilitated by the BCD-based scheme. 

\section{Conclusion}

In this paper, we investigated the fundamental problem of WSR maximization at the IRs in an IRS-assisted SWIPT-MIMO system, subject to satisfying a total weighted harvested power constraint at the ERs. For the formulated non-convex optimization problem, we proposed the PDDAGP algorithm, which is shown to outperform the BCD-based benchmark solution. Numerical results confirmed that the proposed algorithm attains a notably higher WSR, and also increases the operating range of ERs for a given target WSR and target weighted harvested power, compared to the BCD-based benchmark solution. The complexity of the proposed algorithm was shown to be a linear function of the number of IRS elements, while that of the benchmark solution scales with the third power of the  number of reflecting elements of the IRS.

\appendices{}

\section{\label{sec:proofGradX}Proof of Theorem~\ref{thm:gradXClosed}}

Using~\eqref{eq:AugObjDef}, it is straightforward to note that $\nabla_{\mathbf{X}_{m}}\mathcal{R}_{\mu,\rho}\big(\mathbf{X},\boldsymbol{\phi},\tau\big)=\sum_{k\in\mathcal{M}_{\mathrm{I}}}\omega_{k}\nabla_{\mathbf{X}_{m}}R_{k}\big(\mathbf{X},\boldsymbol{\phi}\big)-\big\{\mu+\frac{1}{\rho}f(\mathbf{X},\boldsymbol{\phi},\tau)\big\}\nabla_{\mathbf{X}_{m}}f(\mathbf{X},\boldsymbol{\phi},\tau).$
For the case when $m=k$, using~\eqref{eq:mIR-Rate}, $\nabla_{\mathbf{X}_{m}}R_{k}(\mathbf{X},\boldsymbol{\phi})=\nabla_{\mathbf{X}_{m}}R_{m}(\mathbf{X},\boldsymbol{\phi})$
is given by $\nabla_{\mathbf{X}_{m}}R_{m}(\mathbf{X},\boldsymbol{\phi})=\nabla_{\mathbf{X}_{m}} \big(\ln |\mathbf A_m| - \ln |\mathbf{B}_m| \big)=\nabla_{\mathbf{X}_{m}}\ln|\mathbf{I}+\mathbf{B}_m^{-1/2}\mathbf{Z}_{m}\mathbf{X}_{m}\mathbf{Z}_{m}\herm\mathbf{B}_m^{-1/2}|=\mathbf{Z}_{m}\herm\mathbf{B}_m^{-1/2}\mathbf{C}_{m}^{-1}\mathbf{B}_m^{-1/2}\mathbf{Z}_{m},$
where the last equality follows from~\cite[eqns. (6.195) and (6.200)-(6.207)]{DerivativeBook}, and $\mathbf{C}_{m}\triangleq\mathbf{I}+\mathbf{B}_{m}^{-1/2}\mathbf{Z}_{m}\mathbf{X}_{m}\mathbf{Z}_{m}\herm\mathbf{B}_{m}^{-1/2}$.
Similarly for the case when $m\neq k$, we have $\nabla_{\mathbf{X}_{m}}R_{k}(\mathbf{X},\boldsymbol{\phi})=\nabla_{\mathbf{X}_{m}} \big(\ln |\mathbf A_k| - \ln |\mathbf B_k|\big)=\nabla_{\mathbf{X}_{m}}\ln|\mathbf{I}+\bar{\mathbf{B}}_{k,m}^{-1/2}\mathbf{Z}_{k}\mathbf{X}_{m}\mathbf{Z}_{k}\herm\bar{\mathbf{B}}_{k,m}^{-1/2}|-\nabla_{\mathbf{X}_{m}}\ln|\mathbf{I}+\hat{\mathbf{B}}_{k,m}^{-1/2}\mathbf{Z}_{k}\mathbf{X}_{m}\mathbf{Z}_{k}\herm\hat{\mathbf{B}}_{k,m}^{-1/2}|=\mathbf{Z}_{k}\herm\bar{\mathbf{B}}_{k,m}^{-1/2}\bar{\mathbf{C}}_{k,m}^{-1}\bar{\mathbf{B}}_{k,m}^{-1/2}\mathbf{Z}_{k}\!-\!\mathbf{Z}_{k}\herm\hat{\mathbf{B}}_{k,m}^{-1/2}\hat{\mathbf{C}}_{k,m}^{-1}\hat{\mathbf{B}}_{k,m}^{-1/2}\mathbf{Z}_{k},$
where $\bar{\mathbf{B}}_{k,m}\triangleq\mathbf{I}+\sum_{\imath\in\mathcal{M}_{\mathrm{I}}\setminus\{m\}}\mathbf{Z}_{k}\mathbf{X}_{\imath}\mathbf{Z}_{k}\herm$,
$\bar{\mathbf{C}}_{k,m}\triangleq\mathbf{I}+\bar{\mathbf{B}}_{k,m}^{-1/2}\mathbf{Z}_{k}\mathbf{X}_{m}\mathbf{Z}_{k}\herm\bar{\mathbf{B}}_{k,m}^{-1/2}$,
$\hat{\mathbf{B}}_{k,m}\triangleq\mathbf{I}+\sum\nolimits _{\jmath\in\mathcal{M}_{\mathrm{I}}\setminus\{k,m\}}\mathbf{Z}_{k}\mathbf{X}_{\jmath}\mathbf{Z}_{k}\herm$,
$\hat{\mathbf{C}}_{k,m}\triangleq\mathbf{I}+\hat{\mathbf{B}}_{k,m}^{-1/2}\mathbf{Z}_{k}\mathbf{X}_{m}\mathbf{Z}_{k}\herm\hat{\mathbf{B}}_{k,m}^{-1/2}$.
Following a similar line of argument, $\nabla_{\mathbf{X}_{m}}f(\mathbf{X},\boldsymbol{\phi},\tau)=-\nabla_{\mathbf{X}_{m}}P_{\mathrm{H}}(\mathbf{X},\boldsymbol{\phi})=-(\eta/\tilde{P}_{\mathrm{th}})\sum\nolimits _{\ell\in\mathcal{M}_{\mathrm{E}}}\alpha_{\ell}\boldsymbol{\Xi}_{\ell}\herm\boldsymbol{\Xi}_{\ell}.$
With the help of the derived closed-form expression for $\nabla_{\mathbf{X}_{m}}\mathcal{R}_{\mu,\rho}\big(\mathbf{X},\boldsymbol{\phi},\tau\big)$
and $\nabla_{\mathbf{X}_{m}}f(\mathbf{X},\boldsymbol{\phi},\tau)$,
we obtain the closed-form expression for $\nabla_{\mathbf{X}_{m}}\mathcal{R}_{\mu,\rho}\big(\mathbf{X},\boldsymbol{\phi},\tau\big)$
as given in \emph{Theorem~\ref{thm:gradXClosed}}. This concludes the proof.

\section{\label{sec:proofGradTheta}Proof of Theorem~\ref{thm:gradThetaClosed}}

Using~\eqref{eq:AugObjDef}, it can be noted that $\nabla_{\boldsymbol{\phi}}\mathcal{R}_{\mu,\rho}(\mathbf{X},\boldsymbol{\phi},\tau)=\sum\nolimits _{m\in\mathcal{M}_{\mathrm{I}}}\omega_{m}\nabla_{\boldsymbol{\phi}}R_{m}\big(\mathbf{X},\boldsymbol{\phi}\big)+\big\{\mu+\frac{1}{\rho}f(\mathbf{X},\boldsymbol{\phi},\tau)\big\}\nabla_{\boldsymbol{\phi}}f\big(\mathbf{X},\boldsymbol{\phi},\tau\big).$
Next, to obtain $\nabla_{\boldsymbol{\phi}}R_{m}\big(\mathbf{X},\boldsymbol{\phi}\big)$,
we first use $\nabla_{\boldsymbol{\phi}}\big(R_{m}\big(\mathbf{X},\boldsymbol{\phi}\big)\big)=\nabla_{\boldsymbol{\phi}}\big(\ln|\mathbf{A}_{m}|\big)-\nabla_{\boldsymbol{\phi}}\big(\ln|\mathbf{B}_{m}|\big)$.
Next, we have $\nabla_{\boldsymbol{\phi}}\ln|\mathbf{A}_{m}|=\tr\big\{\mathbf{A}_{m}^{-1}\sum\nolimits _{k\in\mathcal{M_{\mathrm{I}}}}\mathbf{Z}_{m}\mathbf{X}_{k}\nabla_{\boldsymbol{\phi}}\big(\mathbf{Z}_{m}\herm\big)\big\}=\sum_{k\in\mathcal{M}_{\mathrm{I}}}\tr\big\{\mathbf{G}_{m\mathrm{I}}\herm\mathbf{A}_{m}^{-1}\mathbf{Z}_{m}\mathbf{X}_{k}\mathbf{H}_{\mathrm{S}}\herm\mathrm{\nabla_{\boldsymbol{\phi}}}\big(\boldsymbol{\Phi}\herm\big)\big\}$.
Similarly, we can obtain $\nabla_{\boldsymbol{\phi}}\ln|\mathbf{B}_{m}|=\tr\big\{\mathbf{G}_{m\mathrm{I}}\herm\mathbf{B}_{m}^{-1}\mathbf{Z}_{m}\mathbf{X}_{\jmath}\mathbf{H}_{\mathrm{S}}\herm\nabla_{\boldsymbol{\phi}}\big(\boldsymbol{\Phi}\herm\big)\big\}$.
Using the definition of the complex-valued gradient,~\cite[eqn.~(6.153)]{DerivativeBook},
together with the preceding expressions yields $\nabla_{\boldsymbol{\phi}}R_{m}\big(\mathbf{X},\boldsymbol{\phi}\big)=\vecd\big\{\mathbf{G}_{m\mathrm{I}}\herm\mathbf{D}_{m}\mathbf{H}_{\mathrm{S}}\herm\big\},$
where $\mathbf{D}_{m}\triangleq\mathbf{A}_{m}^{-1}\mathbf{Z}_{m}\boldsymbol{\Sigma}-\mathbf{B}_{m}^{-1}\mathbf{Z}_{m}\boldsymbol{\Sigma}_{m}$. Analogously, it can be shown that $\nabla_{\boldsymbol{\phi}}f(\mathbf{X},\boldsymbol{\phi},\tau)=-(\eta/\tilde{P}_{\mathrm{th}})\sum\nolimits _{\ell\in\mathcal{M}_{\mathrm{E}}}\alpha_{\ell}\vecd(\mathbf{G}_{\mathrm{\ell}\mathrm{E}}\herm\boldsymbol{\Xi}_{\ell}\boldsymbol{\Sigma}\mathbf{H}_{\mathrm{S}}\herm).$
With the help of these arguments, we obtain the closed-form expression for $\nabla_{\boldsymbol{\phi}}\mathcal{R}_{\mu,\rho}(\mathbf{X},\boldsymbol{\phi},\tau)$ as given in \emph{Theorem~\ref{thm:gradThetaClosed}}. This completes the proof. 

\bibliographystyle{IEEEtran}
\bibliography{IRS-SWIPT-MIMO}

\end{document}